\documentclass[a4paper,12pt]{article}

\usepackage{amsmath}
\numberwithin{equation}{section}
\usepackage{amsfonts}
\usepackage{color}

\begin{document}

\begin{titlepage}

\begin{flushright}
SNUTP-14-007
\end{flushright}

\vspace*{1.0cm}
\begin{center}
{\Large\bf Momentum bispinor, two-qubit entanglement

\vspace*{0.3cm}

and twistor space}
\end{center}

\vspace*{1.0cm}


\centerline{
Seungbeom Chin$^1$, Sangmin Lee$^{1,2,3}$}

\vspace*{0.5cm}

\begin{center}
{\sl
$^1$ 
School of Physics and Astronomy, Seoul National University, 
Seoul 151-747, Korea
{}\\
$^2$ 
Center for Theoretical Physics, 
Seoul National University, 
Seoul 151-747, Korea
{}\\
$^3$ 
College of Liberal Studies, 
Seoul National University, 
Seoul 151-742, Korea
{}\\
}
\end{center}

\vspace*{2cm}

\begin{abstract}
We re-examine the symmetry structure of massive momentum bispinors in four dimensional Minkowski spacetime and apply the result to the geometry of a two-qubit entanglement system. 
The geometry of entanglement is recovered by restricting the momentum to unit-energy hyperplane inside the future light-cone. 
On a more formal side, to understand Czachor's alternative normalization of the bispinor and its projection property, we analyze the geometry of momentum twistor space. An interesting correpondence between a submanifold of the momentum twistor space and tachyonic particle is also investigated.
\end{abstract}
\end{titlepage}

\section{Introduction}
It is well known that a massless momentum in Minkowski space can be expressed 
as a bispinor, the product of a two-component spinor and its complex conjugate. Similarly, a massive momentum can be written as a sum of two or more bispinors. The bispinor notation 
gives a clear intuition on the action of the spin group SL(2,$\mathbb{C}$). It has found several interesting application in physics. A notable example is the computation of 
scattering amplitudes in gauge field theories.
The bispinor notation is also crucial in the twistor construction \cite{penrose, ward}.

The bispinor decomposition of a massive momentum is known to have an internal symmetry \cite{hughston} which is unitary. According to an argument in ref.~\cite{hughston}, a massive momentum bispinor can be expressed with two independent twistor coordinates. But, Witten \cite{witten} showed that the spinor decomposition can be derived from the momentum bispinor itself, i.e., they are not necessarily understood as some part of twistor spaces. In the literature just mentioned, the investigation of the internal symmetries of a massive bispinor seems to remain incomplete.

In the present paper, we delve deeper into the internal symmetries of the massive momentum bispinor and apply the result to the geometry of a 2-qubit entanglement system. We investigate not only momenta with real mass and positive energy but also momenta with imaginary mass or negative energy. By doing so, we cover the full momentum Minkowski space in a unified manner. We also consider an alternative normalization of the spinors with respect to the mass parameter, first introduced by Czachor \cite{czachor97, czachor}. This normalization and its projection property are reminiscent of the incidence relation of a twistor space, 
and lead us to relate them to the momentum twistor space, i.e., a twistor space in which one side of double fibration is given by a momentum space instead of the position space. 
In either of the two normalizations, the momentum bispinor can be mapped to the reduced density matrices of a 2-qubit entanglement system. The study of this system from a geometric viewpoint was initiated by Mosseri and Dandoloff \cite{mosseri}, and developed further by L\'evay \cite{levay}. We give a slightly new and perhaps more intuitive perspective on the geometry by embedding the system inside the future-pointing null cone of momentum Minkowski space. 

This paper is organized as follows. In section \ref{sec:bispinor}, we analyze all the symmetries underlying the momentum bispinor by using a matrix representation for the bispinor and the notion of coset involving external and internal symmetries. In section \ref{sec:qubit}, we show that the momentum bispinor can be reinterpreted as one of the two reduced density matrices of a 2-qubit system. The subspace of the momentum Minkowski space relevant for the entanglement system is the unit-energy hyperplane inside the future-pointing null cone. The results are compared with L\'evay's  geometric analysis of the 2-qubit system \cite{levay}. In section \ref{sec:twistor}, we revisit Czachor's alternative normalization of the momentum bispinor. 
The spinor set that constitutes the momentum bispinor is understood from the viewpoint of the momentum twistor space, first introduced by Hodges \cite{hodges} to understand dual conformal symmetry of scattering amplitudes of gauge field theories. Here we concentrate on the more intrinsic and fundamental geometric properties of the twistor space, and an interesting correpondence between a submanifold of the momentum twistor space and tachyonic particle is investigated.
Section \ref{sec:conclusion} contains some concluding remarks.

\section{Massive momentum bispinor and its symmetries \label{sec:bispinor}}

It is well known that a massive momentum bispinor matrix can be decomposed into two (or more) massless momentum bispinors \cite{hughston, witten}. In complexified four-dimensional Minkowski space, 
the momentum bispinor can be written as
\begin{align}
p_{\alpha \dot{\beta}} = \pi_{\alpha} \tilde{\pi}_{\dot{\beta}} + \omega_{\alpha} \tilde{\omega}_{\dot{\beta}}\,,
\label{pi-omega}
\end{align}
where $\alpha$ ($\alpha=1,2$) is the spinor index transforming as (1/2,0) under SL$(2,\mathbb{C}$)$\times$SL$(2,\mathbb{C}$) and $\dot{\beta}$ ($\dot{\beta}=1,2$) is the index transforming as (0,1/2). Our convention is 
such that the SL$(2,\mathbb{C})$-invariant tensors are
\begin{align}
\epsilon^{\alpha\beta}= 
\begin{pmatrix} 
0&1\\-1&0
\end{pmatrix}
= \epsilon^{\dot{\alpha}\dot{\beta}},
\qquad
\epsilon_{\alpha\beta}= 
\begin{pmatrix} 
0&-1\\1&0
\end{pmatrix} 
= \epsilon_{\dot{\alpha}\dot{\beta}} \,,
\end{align}
and the matices relating vectors to bispinors ($p_{\alpha\dot{\beta}} = p_\mu (\sigma^\mu)_{\alpha\dot{\beta}}$) are
\begin{align}
(\sigma^{\mu})_{\alpha\dot{\beta}}=(1,\vec{\sigma})_{\alpha\dot{\beta}}, \qquad (\bar{\sigma}^{\mu})^{\dot{\alpha}\beta}=(-1, \vec{\sigma})^{\dot{\alpha}\beta},
\label{pauli4}
\end{align}
with $\vec{\sigma}$ being the standard Pauli matrices.

If we impose the reality condition, $(p_{\alpha \dot{\beta}})^*= p_{\beta \dot{\alpha}}$, the momentum is 
restricted to
\begin{align}
p_{\alpha \dot{\beta}} = \pm ( \pi_{\alpha} \bar{\pi}_{\dot{\beta}} \pm \omega_{\alpha} \bar{\omega}_{\dot{\beta}} ),
\label{real-p}
\end{align}
where $\bar{\pi}_{\dot{\beta}}$ and $\bar{\omega}_{\dot{\beta}}$ are the complex conjugates of $\pi_{\beta}$ and $\omega _{\beta}$. 
The mass squared is 
\begin{align}
m^2 = -p^\mu p_\mu = \textrm{det}(p)= -\frac{1}{2}p_{\alpha\dot{\beta}}p^{\dot{\beta}\alpha}= \pm(\pi_{\alpha}\omega^{\alpha})(\bar{\pi}_{\dot{\alpha}}\bar{\omega}^{\dot{\alpha}})\equiv
 \pm (\pi \cdot \omega) (\bar{\pi} \cdot \bar{\omega})\,.
\label{mass-square}
\end{align}
The sign of $m^2$ is the same as the relative sign between the two terms 
in \eqref{real-p}. In the case of real mass ($m^2>0$), the overall sign in \eqref{real-p} gives the sign of $p_0$, which 
is $(-1)$ times the energy.  
In the case of imaginary mass ($m^2<0$), the overall sign is immaterial,
as any two space-like vectors are connected by an SO$(1,3)$ transformation.  
In terms of the spinors, the sign flip amounts to a symmetry transformation exchanging $\pi_\alpha$ and 
$\omega_\alpha$.

The representation \eqref{real-p} has some redundancy; different 
values of $\pi_\alpha$ and $\omega_\alpha$ may give the same 
$p_{\alpha\dot{\beta}}$. In the rest of this section, we analyze this ``gauge" (or internal) symmetry \cite{penrose, hughston} for the real and imaginary mass cases separately. The results will be used in the subsequent discussions.

\subsection{Momentum with real mass \label{sec:real-mass}}

One way to understand the gauge symmetry is 
to combine $\pi$, $\omega$ into a 
$2 \times 2$ matrix, 
\begin{align}
A \equiv 
\begin{pmatrix} 
\pi_{1} & \pi_{2} 
\\
\omega_{1} & \omega_{2} 
\end{pmatrix} \,.
\label{A-matrix}
\end{align}
We immediately see that the momentum matrix,
\begin{align}
p =\pm ( A^{\dagger}A )^{T} \,,
\end{align}
is invariant under the U$(2)$-left action on $A$, 
\begin{align}
A \rightarrow UA\,, \quad U \in \mathrm{U}(2)\,.
\end{align}

The gauge symmetry removes $\mathrm{dim}[\mathrm{U}(2)]=4$ components 
from the eight (four complex) components of the matrix $A$.  
This agrees with the fact that a momentum has four components 
before imposing the on-shell condition $-p^\mu p_\mu = m^2$.
To account for the on-shell condition, we first define 
the complexified mass $\mu$ as
\begin{align}
\mu = m e^{i\psi} \equiv \textrm{det}(A) = (\pi \cdot \omega)  \,.
\end{align}
The case $\mu=0$ should be treated separately. We refer the readers to appendix \ref{sec:app} for details. For $\mu \neq 0$, we can extract the mass-dependence from $A$ by setting 
\begin{align}
A = \sqrt{\mu} A' \qquad (\textrm{det}A'= 1) \,.
\label{sl2c}
\end{align}
The diagonal U$(1)$ subgroup of the U$(2)$ can be used to remove the angle $\psi$.  
By definition, $A' \in \mathrm{SL}(2,\mathbb{C})$ 
and the remaining gauge symmetry is SU$(2)$. 
Thus we recover the well-known result that 
the space of momentum with a fixed mass 
is the coset, 
\begin{align}
\mathrm{SL}(2,\mathbb{C})/\mathrm{SU}(2) \simeq 
\mathrm{SO}(1,3)/\mathrm{SO}(3) \,.
\end{align}

For concrete computations, it is necessary to ``fix the gauge", i.e., decompose $A'$ into a gauge action and a unique representative of the coset. For instance, the gauge choice of ref.~\cite{aulbach} 
takes the form  
\begin{align}
A' = U'
\begin{pmatrix}
\frac{c}{\lambda} & \frac{1}{\lambda}
\\
-\lambda & 0 
\end{pmatrix} 
\equiv U' F \,, 
\qquad   U' \in \mathrm{SU(2)}\,,
\label{corep}
\end{align}
where $c$ is a complex number and $\lambda$ is real and positive. 
A short computation relates $c$ and $\lambda$ to the momentum components, 
\begin{align}
p_0 + p_3 = m\left(\frac{|c|^2}{\lambda^2} + \lambda^2 \right)\,, 
\quad
p_0 - p_3 = \frac{m}{\lambda^2} \,,
\quad 
p_1 - i p_2 = \frac{mc}{\lambda^2} \,.
\end{align}
A shift in the phase of $c$ corresponds to a rotation 
in the $(p_1,p_2)$-plane, while the scaling, 
$\lambda \rightarrow \kappa \lambda$, $c\rightarrow \kappa^2 c$ 
corresponds to a boost in the $(p_0,p_3)$-plane.  
There is an alternative scaling $\lambda \rightarrow \kappa' \lambda$ with $c$ fixed, 
which rescales the spinors as $\pi \rightarrow \kappa\pi$ and $\omega \rightarrow \omega/\kappa'$. It will play a role in section \ref{sec:twistor}.

Finally, to facilitate the analysis of section 3, 
we write down the metric on the momentum Minkowski space. 
In the standard hyperbolic coordinate,
\begin{align}
(p_{0}, p_{1}+ip_{2},p_{3})=m(\pm\cosh{\zeta},\sinh {\zeta}\sin\theta e^{i\phi},\sinh{\zeta}\cos\theta)\,,
\label{hyperboloid}
\end{align}  
the metric takes the form 
\begin{align}
ds^{2} = dp^\mu dp_\mu 
=- dm^{2}+m^{2}\left[ d\zeta^{2}+\sinh^{2}{\zeta}(d\theta^{2}+\sin^{2}\theta d\phi^{2}) \right].
\end{align}
The relation between ($\zeta,\theta, \phi$) and ($\lambda, C$) is as follows:
\begin{align}
c=\frac{\sinh{\zeta}\sin\theta}{\cosh{\zeta}-\sinh{\zeta}\cos\theta}e^{i\phi}\,,
\quad
\lambda=\frac{1}{(\cosh{\zeta}-\sinh{\zeta}\cos\theta)^{\frac{1}{2}}}.
\end{align} 
This metric coincides with the standard coset metric of GL$(2,\mathbb{C})$/U$(2)$. 
Similarly, the metric within the square bracket coincides 
with the coset metric of SL$(2,\mathbb{C})$/SU$(2)$. 
In terms of the coset representative $F$ of SL$(2,\mathbb{C})$/SU$(2)$ defined in \eqref{corep}, the explicit coset construction of the metric proceeds as follows:
\begin{align}
& dF F^{-1} = \frac{dc}{2\lambda^2}(\sigma_1 + i \sigma_2) -\frac{d\lambda}{\lambda} \sigma_3 \equiv (\vec{e} + i \vec{f}) \cdot \frac{\vec{\sigma}}{2} 
\nonumber \\
&
\quad
\Longrightarrow  
\quad 
ds^2_{\mathrm{SL}(2,\mathbb{C})/\mathrm{SU}(2)} = \vec{e}\cdot\vec{e} = \frac{|dc|^2}{\lambda^4} + 4 \frac{d\lambda^2}{\lambda^2} \,.
\label{co-met}
\end{align}

\subsection{Momentum with imaginary mass}

In this case the complexified mass is expressed as
\begin{align}
\mu= (\pi \cdot \omega)= |m| e^{i\psi}
\end{align}
 with $m=i|m|$. The momentum matrix is given by
\begin{align}
p = ( A^{\dagger}\sigma_{3}A)^{T},
\end{align}
which is invariant under the U$(1,1)$-left action on $A$,
\begin{align}
A \rightarrow VA, \qquad V\in \textrm{U}(1,1) \,.   
\end{align}

We separate the mass dependence from $A$ as we did in section \ref{sec:real-mass},
\begin{align}
A=\sqrt{\mu}A' \qquad (\textrm{det}A'=1).
\end{align}
The moduli space of space-like momenta at a fixed $k\equiv |m|$ is identified with the coset,
\begin{align}
\mathrm{SL}(2,\mathbb{C})/\mathrm{SU}(1,1) 
\simeq 
\mathrm{SO}(1,3)/\mathrm{SO}(1,2) \,.
\end{align}
We choose the gauge for the coset representative of $A'$ as
\begin{align}
A'= 
V'             
\begin{pmatrix} e^{\zeta/2} \cos\frac{\theta}{2} & e^{\zeta/2+i\phi}\sin\frac{\theta}{2} \\
                          - e^{-\zeta/2-i\phi}\sin\frac{\theta}{2} & e^{-\zeta/2}\cos\frac{\theta}{2}
\end{pmatrix}\,,
 \qquad   V' \in \mathrm{SU(1,1)}\,.
\end{align}
It is equivalent to the following parametrization of the momenta,
\begin{align}
(p_{0}, p_{1}+ip_{2},p_{3})=k(\sinh{\zeta} ,\cosh{\zeta}\sin\theta \,e^{i\phi}, \cosh{\zeta}\cos\theta) \,.
\end{align}
The metric in this coordinate system reads 
\begin{align}
ds^{2} &= dp^\mu dp_\mu = dk^{2}+k^{2}[-d\zeta^{2}+\cosh^{2}{\zeta}(d\theta^{2}+\sin^{2}\theta d\phi^{2})]\,.
\end{align}

\section{Momentum/2-qubit correspondence \label{sec:qubit}}

In this section, we will examine the formal similarity between the real-mass momentum matrix with positive energy and the reduced density matrix of entangled bipartite qubits. 
We first summarize some well-known results concerning the 2-qubit pure state entanglement, and relate them to the properties of the real-mass momentum bispinor. The resulting correspondence offers a slightly new perspective on the 
geometric analysis of the 2-qubit entanglement system initiated by \cite{mosseri} and further developed by \cite{levay}.

\subsection{Reduced density matrix as a momentum bispinor}

We begin with writing the two-particle state vector as
\begin{align}
|\Psi\rangle= \frac{1}{\sqrt{2}}\sum_{i, \alpha=1,2}A_{i\alpha} \Big( |i\rangle_{A} \otimes |\alpha\rangle_{B} \Big),
\label{A-quantum}
\end{align}
where $A_{i \alpha}$ is considered as the elements of a $2 \times 2$ matrix $A$. The total (pure) density matrix is given by 
$\rho =|\Psi \rangle \langle \Psi |$.
By taking the partial trace over each of the two subsystems, we obtain two reduced density matrices given by
\begin{align}
\rho_{A}=\frac{1}{2} AA^{\dagger}, \qquad \rho_{B} =\frac{1}{2} (A^{\dagger}A)^{T}.
\end{align}

When the two subsystems are maximally entangled, the density matrices become $\rho_{A} =\rho_{B}=\frac{1}{2}I_{2}$. So we have $A \in$ U(2) and the von Neumann entropy $S=\log_{2}2=1$. When two subsystems are separated, $S$ vanishes and $\mathrm{det}(A)=0$.

We partition the given Hilbert space $\mathbb{C}^{2} \times \mathbb{C}^{2}$ into equivalent classes to categorize the different degree of entanglement. Two states are defined to be LOCC (local operation and classical communication)-equivalent 
when they are connected by an SU(2)$\times$SU(2) rotation. In other words, two given states $|\Psi \rangle$ and $|\Psi' \rangle$ are LOCC-equivalent iff
\begin{align}
|\Psi' \rangle = (U_{A}\otimes U_{B})|\Psi \rangle \,, \qquad U_{A},U_{B} \in SU(2) \,.
\end{align}

Now, let us substitute \eqref{A-matrix} into \eqref{A-quantum} and rewrite \eqref{A-quantum} as 
\begin{align}
|\Psi \rangle  =\frac{1}{\sqrt{2}}\Big( \pi_{1}|11\rangle +\pi_{2}|12\rangle +\omega_{1}|21\rangle + \omega_{2}|22\rangle  \Big) \,,
\end{align}
where $|i\alpha\rangle$ denotes $|i\rangle_{A} \otimes |\alpha\rangle_{B}$.  
The explicit form of the density matrices are 
\begin{align}
\rho_{A}= \frac{1}{2}
\begin{pmatrix} 
\pi_{1}\bar{\pi}_{\dot{1}} +  \pi_{2}\bar{\pi}_{\dot{2}}& \pi_{1}\bar{\omega}_{\dot{1}}+\pi_{2}\bar{\omega}_{\dot{2}} \\
\omega_{1}\bar{\pi}_{\dot{1}} + \omega_{2}\bar{\pi}_{\dot{2}} & \omega_{1}\bar{\omega}_{\dot{1}} +\omega_{2}\bar{\omega}_{\dot{2}} \end{pmatrix}
\,,
\quad
(\rho_{B})_{\alpha\dot{\beta}} = \frac{1}{2}p_{\alpha\dot{\beta}}
\label{rho-A}
\end{align}
with $p_{\alpha\dot{\beta}}$ defined in \eqref{real-p}. 
Note that the density matrix satisfies the same Hermiticity condition as the momentum bispinor for real Minkowski space; see \eqref{pauli4}.
The normalization condition for the state $|\Psi\rangle$ implies
\begin{align}
\langle \Psi |\Psi \rangle  =\mathrm{tr}\rho_{A} =\mathrm{tr}\rho_{B}=p_{0} = 1.
\end{align}

\begin{figure}
\setlength{\unitlength}{5cm}
\begin{center}
\begin{picture}(2.5,1)
\put(1,0.72){null cone}
\put(0.53,0.95){$p_{0}$}
\put(0,0.59){\line(1,0){1}}
\put(0.5,0.3){\vector(1,1){0.5}}
\put(0.92,0.18){$+|\vec{p}|$}
\put(-0.1,0.18){$-|\vec{p}|$}
\put(0.95,0.52){$p_{0}=1$}
\put(0.5,0.3){\vector(-1,1){0.5}}
\put(0.5,0.1){\vector(0,1){0.9}}
\put(0.5,0.3){\vector(1,0){0.5}}
\put(0.5,0.3){\vector(-1,0){0.5}}
\qbezier(0.01,0.81)(0.25,0.525)(0.5,0.5)
\qbezier(0.5,0.5)(0.75,0.525)(0.99,0.81)
\qbezier(0.01,0.81)(0.25,0.58)(0.5,0.59)
\qbezier(0.5,0.59)(0.75,0.58)(0.99,0.81)
\put(0.45,-0.1){(a)}
\put(2,0.5){\circle*{0.01}}
\put(2,0.5){\circle{0.4}}
\put(2,0.5){\circle{0.9}}
\put(2.5,0.5){$m=0$}
\put(1.9,0.53){$m=1$}
\put(1.8,0.23){$0<m<1$}
\put(1.94,-0.1){(b)}
\end{picture}
\end{center}
\caption{Intersection loci between mass-shell condition and unit energy condition.}
\label{fig:p-ball}
\end{figure}
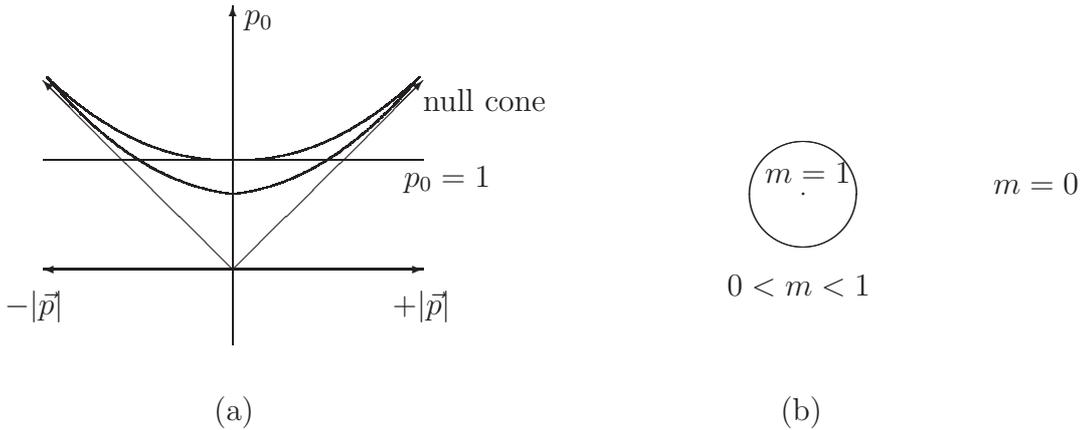

Since we are dealing with real-mass momenta, the mass must not be bigger than 1 for the state to permit unit energy. So the region that corresponds to the entangled 2-qubit is $p_{0}=1$ hyperplane inside the null cone (Fig.~\ref{fig:p-ball}a). 
In this context, $m=|\mathrm{det}A|$ is identified with the concurrence \cite{wootters1} in the quantum information literature.\footnote{See \cite{wootters2} for general 
definition and properties of concurrence.}
Different $m$ ($0  \le m \leq 1$) gives a different value of entanglement measure, i.e., entropy in the bipartite case. The von Neumann entropy depends on $m$ as 
\begin{align}
S=-(\lambda_{+}\log_{2}{\lambda_{+}} +\lambda_{-}\log_{2}{\lambda_{-}})\,,
\quad
\lambda_{\pm}=\frac{1}{2} (1 \pm \sqrt{1-m^{2}}).
\end{align}

 Qubits are maximally entangled when $m=1$ (the upper curve inside the null cone of Fig.~\ref{fig:p-ball}a). The intersection locus of this curve and the $p_{0}=1$ plane consists of a single point in the Minkowski space, namely, 
$p_{\mu}=(1,0,0,0)$, 
and we have $A \in$ U(2), $p =I_{2}$, and $S=1$. As the mass decreases over the range, $0< m<1$ (the lower curve in Fig.~\ref{fig:p-ball}a), the entropy decreases as well but never reaches zero. So the quantum state is still entangled. The intersection locus constitutes an $S^{2}$ (the sandwiched circle in Fig.~\ref{fig:p-ball}b). Finally, when $m=0$ the entropy vanishes and the quantum state is separable. See appendix \ref{sec:app} for the geometry of the space of the separable states. 

LOCC-equivalent classes of 2-qubit states are represented clearly in the momentum Minkowski space. In the momentum representation, one of the SU(2)$\times$SU(2) is a subgroup of the internal U(2) symmetry and the other is the spatial rotation in the Minkowski space. So all the points in the intersection of the $p_{0}=1$ plane and an equal mass curve are LOCC-equivalent to each other (Fig.~\ref{fig:p-ball}b) if they have the same internal phase of U(1) $\subset$ U(2) that are hidden in Fig.~\ref{fig:p-ball}. If we reinstate the U(1) phase, the submanifold 
of equal entropy is $S^{2} \times S^{1}$ for $0<m<1$ and $S^2$ for $m=0$.

\subsection{Geometry of 2-qubit entanglement revisited}

A geometric approach to the 2-qubit entanglement system was initiated by Mosseri and Dandoloff \cite{mosseri}. They used the Hopf fibration 
of $S^{7}$ with $S^3=\mathrm{SU}(2)$ fibered over the base $S^4$. They showed that the base $S^{4}$ is {\em entanglement sensitive}, i.e., it 
can be divided into submanifolds of fixed entanglement.  
L\'evay \cite{levay} further developed this approach, 
clarified the coordinates on the submanifolds 
and discussed metrics and connections. 
In this subsection, we show that some of the main results 
of \cite{mosseri, levay} can be understood intuitively in 
terms of the momentum bispinor without invoking 
the full machinery of quanternions. 
 
To begin with, recall that the normalization condition of the state $|\Psi\rangle$ is given by
\begin{align}
|\pi_{1}|^{2} +|\pi_{2}|^{2} +|\omega_{1}|^{2} +|\omega_{2}|^{2} 
=2p_{0}= 2,
\end{align}
which restricts $\mathbb{C}^{2} \times \mathbb{C}^{2}$ to $S^{7}$. 
L\'evay \cite{levay} showed that the Hopf map from $S^7$ to the base $S^{4}$ 
can be described by one of the two density matrices (3 variables) and det$A$ (2 variables) subject to a constraint. In our notation, 
the constraint is simply   
\begin{align}
&\mathrm{det}(2\rho) = \mathrm{det}(p)= p_{0}^{2}-|\vec{p}|^{2}=1-|\vec{p}|^{2} = |\mu|^2 = |\mathrm{det}A|^2 \,.
\label{S4-1}
\end{align}
Reparametrizing the complex mass as $\mu  = \mu_1+i\mu_2$ $(\mu_1,\mu_2 \in \mathbb{R})$, we have 
\begin{align}
p_{1}^{2}+p_{2}^{2}+p_{3}^{3}+\mu_1^2 + \mu_2^2 = 1  \,,
\label{S4-2}
\end{align}
which defines a unit $S^4$ in $\mathbb{R}^5$.  
Recall that the internal SU(2)$_{L}$ symmetry, which acts on the observer $A$, leaves all variables in \eqref{S4-2} invariant. Thus SU(2)$_{R}$ is 
naturally identified with the isometry of the fiber $S^3$. The other SU(2)$_R$ symmetry, which acts on the observer $B$, act on $\vec{p}$ as the spatial rotation and suggests a natural way to divide $S^4$ into subspaces. 
Let us rewrite \eqref{S4-1} as
\begin{align}
 p_{1}^{2}+p_{2}^{2}+p_{3}^{2}=1-m^{2},
\end{align}
Since $m^{2}$ is restricted to the range $0 \le m^{2} \leq 1$, 
the 3-vector $\vec{p}$ spans a 3-ball $B^{3}$. 
At each point in the interior of $B^3$ ($m^2<1$), the phase $\psi$ of $\mu=m e^{i\psi}$ spans a circle. On the boundary of $B^3$ ($m^2=0$), 
which corresponds to separable states, 
the circle shrinks to a point. Hence we recover \cite{mosseri,levay}   
\begin{align}
S^{4} \setminus S^2 \simeq \mathrm{Int}(B^{3}) \times S^{1},
\label{S4-conformal}
\end{align}
where $\mathrm{Int}(B^{3})$ is the interior of a unit 3-ball, 
and the $S^2$ on the left-hand side represent 
the separable states. 
To write the metric on the $S^4$,  
we use the coordinate defined in \eqref{hyperboloid} 
and 
restrict it to the slice $p_{0}=m\cosh{\zeta}=1$,
\begin{align}
ds^{2} = |d\vec{p}|^2 + |d\mu|^2 
=\frac{1}{\cosh^{2}{\zeta}}\Big[d\zeta^{2}+\sinh^{2}{\zeta}(d\theta^{2}+\sin^{2}\theta d\phi^{2}) +d\psi^{2} \Big].
\end{align}
Aside from the overall conformal factor, the metric describes $\mathrm{Int}(B^3)\times S^1$. 
A slight novelty in our route to obtain this metric is that the change of variable \eqref{hyperboloid} follows naturally from the geometry of real-mass momentum space, while in ref.~\cite{levay} the substitution of $\tanh^{2}X$ into $R^{2}\equiv 1-|\textrm{det}A|^{2}$ is merely based on the range of $R$. 

Table \ref{table:dic} summarizes how various quantities are mapped between the 2-qubit system and the momentum bispinor. 

\begin{table}[htbp]
	\begin{center}
    \begin{tabular}{| l | l |}
    \hline
    2-qubit & momentum bispinor \\ \hline \hline
    concurrence $|$det$A|$ & mass $m$  \\ \hline
    det$A$ & $\mu = m e^{i\psi}$ \\ \hline
    gauge SU(2)$\times$SU(2) & SU(2)$_{R{\rm (gauge)}}$ $\times$ SU(2)$_{L{\rm (rotation)}}$  \\  \hline
    $\mathrm{tr}\rho=\mathrm{tr}\rho_{A}=\mathrm{tr}\rho_{B}=1$ & $p_{0}=1$ (unit energy) \\ \hline
    $S^{4} \simeq$ $B^{3} \times S^{1}/\mathbb{Z}_{2}$ & $|\vec{p}|^{2}+|\mu|^{2}=1$ \\ \hline
    seperable state $\;\; \leftrightarrow\;\;$ $\partial(B^{3})$ & massless momentum \\ \hline   
    \end{tabular}
    \caption{A dictionary between entagled 2-qubit system and momentum bispinor}
    \label{table:dic}
    \end{center}
\end{table}

\newpage

\section{From momentum bispinor to momentum twistor\label{sec:twistor}}

In the relativistic unit with the speed of light set to unity, the 4-momentum has mass dimension 1. The bispinor decomposition of the momentum assign dimension 1/2 to the spinors. 
There are other ways to normalize the spinors, which renders the spinors dimensionless and extract the mass as an independent variable. For example, Woodhouse \cite{woodhouse} rewrites the massive momentum bispinor as
\begin{align}
p_{\alpha \dot{\beta}} = \pm m ( \hat{\pi}_{\alpha} \bar{\hat{\pi}}_{\dot{\beta}} \pm \hat{\omega}_{\alpha} \bar{\hat{\omega}}_{\dot{\beta}} ),
\end{align}
where the hats on the spinors mean that they are 
normalized to satisfy ($\hat{\pi} \cdot \hat{\omega}) = e^{i\psi}$. 
 
Czachor \cite{czachor97, czachor} proposed yet another normalization that allows for an interesting geometric interpretation to be discussed in this section,
\begin{align}
p_{\alpha \dot{\beta}} = \pm (\hat{\pi}_{\alpha} \bar{\hat{\pi}}_{\dot{\beta}} \pm m^{2}\hat{\omega}_{\alpha} \bar{\hat{\omega}}_{\dot{\beta}}) \,,
\label{czachorp}
\end{align}
with restrictions ($\hat{\pi} \cdot \hat{\omega}) =e^{i\psi}$ and $0\leq m < \infty$. 
\footnote{Refs.~\cite{czachor97, czachor} consider only the case $p_{\alpha \dot{\beta}} =\hat{\pi}_{\alpha} \bar{\hat{\pi}}_{\dot{\beta}} + m^{2}\hat{\omega}_{\alpha} \bar{\hat{\omega}}_{\dot{\beta}} $, with $\hat{\pi} \cdot \hat{\omega} = 1$.  It corresponds to the real-mass and positive-energy case in our discussions with the phase variable fixed.} 
The internal symmetries for this form are basically the same as those for \eqref{pi-omega}, except that 
$(\hat{\pi}, m\hat{\omega})$ constitute a pair for the rotations now, and we define a matrix $C$ as
\begin{align}
C \equiv 
\begin{pmatrix}
\hat{\pi}_{1} & \hat{\pi}_{2}\\
m\hat{\omega}_{1} & m\hat{\omega}_{2}
\end{pmatrix}            
=
\frac{1}{|\textrm{det}A|^{\frac{1}{2}}} 
\begin{pmatrix}
 1 & 0\\
 0 & m
\end{pmatrix} A
\,,
\end{align}
which makes us rewrite the momentum as
\begin{align}
p= \pm (C^{\dagger}C)^{T} \phantom{a} \textrm{(real mass)}, \qquad p=\pm (C^{\dagger}\sigma_{3}C)^{T} \phantom{a} \textrm{(imaginary mass)}. 
\label{czachorpc}
\end{align}
When $m\neq 0$, this map is invertible and all the results in section \ref{sec:bispinor} and \ref{sec:qubit} remain valid. But, as pointed out in \cite{czachor97}, the new normalization \eqref{czachorp} has a minor advantage that we can 
reach the $m=0$ ``boundary" while keeping $(\hat{\pi}\cdot \hat{\omega})$ finite. 

The next thing to note is that the mass dimension of $\omega$ 
has shifted from $(+1/2)$ to $(-1/2)$. We recall that a pair of spinors with opposite mass dimensions constitute the 
building block of the twistor space. Indeed, by multiplying 
\eqref{czachorp} by $\bar{\hat{\omega}}$, we obtain
\begin{align}
i p_{\alpha \dot{\beta}}\bar{\hat{\omega}}^{\dot{\beta}}= \pm i\hat{\pi}_{\alpha} \bar{\hat{\pi}}_{\dot{\beta}} \bar{\hat{\omega}}^{\dot{\beta}}=\hat{\pi}_{\alpha}
\label{p-incidence}
\end{align}
for spinors that satisfy $\bar{\hat{\pi}} \cdot \bar{\hat{\omega}} =\mp i$. This is identical to the standard ``incidence relation" 
in the twistor literature, except that $p_{\alpha\dot{\beta}}$ has 
replaced $x_{\alpha\dot{\beta}}$, the coordinates of the position Minkowski space written in bispinor form. 

We should stress that what looks like the incidence relation 
\eqref{p-incidence} by itself does not guarantee that $(\hat{\pi},\hat{\omega})$ forms a genuine twistor space. 
Our task in this section is to fill up possible gaps and show that $(\hat{\pi},\hat{\omega})$ indeed can be understood as the subset of a twistor space coordinate.
To do so, we use a twistor space that has the momentum bispinor as the two-dimensional flag manifold of flat twistor space, i.e. the momentum twistor space of Hodges \cite{hodges}. Next, we show that Czachor's bispinor form is the solution to the incidence relation in momentum twistor space with reality condition. The momentum bispinor corresponding to a submanifold of the twistor space, ($\pi\cdot\omega)=0$ for nonzero $\pi$ and $\omega$, is also analyzed, which  turns out to be the momentum of classical tachyon. 

\subsection{Geometry of momentum twistor space \label{sec:k-twi}}

{}From a purely geometric point of view, the construction of the double fibration system of the momentum twistor space has a lot in common with that of the position twistor space. So the standard approach to the position twistor geometry, e.g., \cite{penrose, ward}, can be exploited to achieve our present goal. We should stress that the two twistor spaces may have totally different physical applications. For example, in the position twistor space, the norm of twistor coordinates is interpreted as the helicity of a particle, but such an interpretation is unknown for the momentum twistor space considered here. 
In this subsection, we will restrict our attention to the geometric construction of the double fibration system of momentum twistor space.

 A fixed four-complex-dimensional vector space is called the (flat) twistor space, usually denoted by $\mathbb{T}_{4}$. A twistor $Z^{A}$, with components 
\begin{align}
(Z^{0}, Z^{1}, Z^{2}, Z^{3}) \equiv (\pi_{1},\pi_{2}, \bar{\omega}^{\dot{1}}, \bar{\omega}^{\dot{2}}) \in \mathbb{C}^{4}
\end{align}
 is an element of $\mathbb{T}_{4}$. The nondegenerate Hermitian form on $\mathbb{T}_{4}$, denoted by $\Phi$ is preserved under SU$(2,2)$ rotations. So with
\begin{align}
\Phi_{AB}= 
\begin{pmatrix} 
0 & I_{2} \\
I_{2} & 0 
\end{pmatrix}_{AB} ,
\label{phi1}
\end{align}
a twistor and the conjugate twistor which belongs to the dual space $\mathbb{T}^{*}_{4}$ are given by 
\begin{align}
Z^{A}= 
\begin{pmatrix} 
\pi_{\alpha} 
\\
\bar{\omega}^{\dot{\alpha}}
\end{pmatrix}\,, 
\qquad 
\bar{Z}_{A}\equiv (Z^\dagger)^{B} \Phi_{BA} =
(\omega^{\alpha} , \bar{\pi}_{\dot{\alpha}}) \,.
\end{align}
The SU$(2,2)$-invariant norm, 
\begin{align}
|Z|^2 \equiv \bar{Z}_{A} Z^{A} =\pi_{\alpha}\omega^{\alpha} + \bar{\omega}^{\dot{\alpha}} \bar{\pi}_{\dot{\alpha}} \,,
\label{twi-norm}
\end{align}
divides the given twistor space into three domains: $Z^{2} > 0$, $Z^{2}=0$, and $Z^{2} < 0$. 
The standard way to understand the map between the twistor space 
and Minkowski space makes use of flag manifolds. For $\mathbb{C}^n$,  
given a sequence integers, $0<d_{1}< \cdots < d_{m}<n$, a flag manifold is defined to be a collection of ordered sets of vector spaces, 
\begin{align}
F_{d_{1}, \cdots, d_{m}}(\mathbb{C}^n) \equiv \{(V_{1}, \cdots, V_{m}): V_{1} \subset V_{2} \subset \cdots \subset V_{m} \subset \mathbb{C}_n\,, \mathrm{dim}(V_i) = d_i \} \,.
\end{align}
For the twistor space, $\mathbb{T}_4 \cong \mathbb{C}^4$, 
the most relevant flag manifolds are $F_{12}$, $F_{1}$ and $F_{2}$. 
First, $F_1$ is the collection of lines in $\mathbb{C}^4$, or $\mathbb{C}P^{3}$, the complex projective space.  Similarly, $F_{2}$ defines the Grassmannian manifold $G_{2,4}(\mathbb{C})$, which has dimension 4. Upon a suitable gauge fixing, $F_2$ is identified with the (conformally compactified) complexified Minkowski space $\mathbb{M}$. 
Finally, $F_{12}$ can be shown to be 
$F_{12} \cong \mathbb{M} \times \mathbb{C}P_{1}$. 
The local coordinates for the flag manifolds are as follows.
\begin{align}
\nonumber
&F_1 : (\pi_\alpha, \bar{\omega}^{\dot\alpha}) \in \mathbb{C}^4 \sim \lambda
(\pi_\alpha, \bar{\omega}^{\dot\alpha})\,, 
\quad 
\lambda \in \mathbb{C}^*\,,
\\
&F_2 : 
\begin{pmatrix}
ip_{11} & ip_{21} & 1 & 0 \\
ip_{12} & ip_{22} & 0 & 1  
\end{pmatrix} \in G_{2,4}(\mathbb{C}) \,,
\quad p_{\alpha\dot{\beta}} \in \mathbb{M} \,,
\\
&F_{12} : (p_{\alpha \dot{\beta}}, \bar{\omega}^{\dot{\alpha}}) 
\,.
\nonumber
\end{align} 
They admit a double fibration structure, 
\begin{align}
&\quad\quad \; F_{12}
\nonumber \\
&\mu \swarrow \qquad \searrow \nu
\label{dbfb}
\\
&F_{1} \qquad\qquad F_{2}
\nonumber 
\end{align}
The projection $\nu$ simply ``forgets" 
$\bar{\omega}^{\dot{\alpha}}$ in $(p_{\alpha \dot{\beta}}, \bar{\omega}^{\dot{\alpha}})$. 
The projection $\mu$ maps $(p_{\alpha \dot{\beta}}, \bar{\omega}^{\dot{\alpha}}) \in F_{12}$ to $(\pi_{\alpha} = i p_{\alpha \dot{\beta}} \bar{\omega}^{\dot{\beta}} , \bar{\omega}^{\dot{\alpha}}) \in F_1$. The image of this projection is $P^{3} \cong \mathbb{C}P^{3} \setminus \mathbb{C}P^{1}$, where the deleted line corresponds to ($\pi \neq 0, \bar{\omega} =0$).
In view of the double fibration structure, $F_{12}$ is called the correspondence space between $P^{3}$ and $\mathbb{M}$.

\subsection{Momentum bispinor as $F_{2}$ of momentum twistor space \label{sec:f2}}

In this subsection, we will show how 
the general solution of the incidence relation given in section \ref{sec:k-twi} under the reality condition on the momentum Minkowski space is mapped to Czachor's momentum bispinor \eqref{czachorp}.  

We start from the incidence relation
\begin{align}
ip_{\alpha \dot{\beta}}\bar{\omega}^{\dot{\beta}} = \pi_{\alpha}\,.
\label{incidence}
\end{align}
Multiplying both sides by $\omega^\alpha$ and taking the real part, we see that the the necessary and sufficient condition for the momentum bispinor $p_{\alpha\dot{\beta}}$ to be Hermitian is precisely
\begin{align}
\bar{Z}_{A} Z^{A}= \pi_{\alpha}\omega^{\alpha}+\bar{\omega}^{\dot{\alpha}}\bar{\pi}_{\dot{\alpha}}=0,
\label{hermi-1}
\end{align}
if $\pi_\alpha\omega^\alpha \neq 0$. 
The general solution to this constraint is
\begin{align}
\pi_{\alpha}\omega^{\alpha}= \pm r^{2} i \qquad (r \textrm{ is real and not zero}).
\label{hermi-2}
\end{align}
Here $r$ represents the real rescaling of $Z^{A}$. Since the incidence relation \eqref{incidence} is invariant under rescaling 
by any non-zero number, we may work with the rescaled spinors
\begin{align}
\pi^0_\alpha \equiv \frac{1}{r} \pi_\alpha \,,
\quad 
\omega^0_\alpha \equiv \frac{1}{r} \omega_\alpha \,.
\end{align}
As $\pi^0_\alpha$ and $\omega^0_\alpha$ are linearly independent\ 
($\pi_\alpha \omega^\alpha\neq0$), 
any bispinor can be expanded in the following form, 
\begin{align}
p_{\alpha\dot{\beta}} = 
c_1 \pi_\alpha \bar{\pi}_{\dot{\beta}} +
c_2 \pi_\alpha \bar{\omega}_{\dot{\beta}} +
c_3 \omega_\alpha \bar{\pi}_{\dot{\beta}} +
c_4 \omega_\alpha \bar{\omega}_{\dot{\beta}} \,.
\label{generalp}
\end{align}
The Hermiticity condition requires that $c_1^* = c_1$, $c_2^*=c_3$ and $c_4^* = c_4$. The incidence relation implies that 
$c_3=0$, $c_1 = \pm i$. 
Changing the notation from $c_4$ to $\pm m^2$, we can summarize the general solutions as
\begin{align}
&p_{\alpha \dot{\beta}}^{(+)} = \pi^{0}_{\alpha} \bar{\pi}^{0}_{\dot{\beta}} +m^{2} \omega^{0}_{\alpha} \bar{\omega}^{0}_{\dot{\beta}} \qquad (-\infty \leq m^{2} \leq \infty), \qquad (\pi^{0}\cdot\omega^{0}=+i)
\label{+ep}
\\
&p_{\alpha \dot{\beta}}^{(-)} = -(\pi^{0}_{\alpha} \bar{\pi}^{0}_{\dot{\beta}} +m^{2} \omega^{0}_{\alpha} \bar{\omega}^{0}_{\dot{\beta}}) \qquad (-\infty \leq m^{2} \leq \infty). \qquad (\pi^{0}\cdot\omega^{0}=-i)
\label{-ep}
\end{align}
Comparing these results with \eqref{czachorp}, we arrive at the identification ($\pi^{0}, \omega^{0})=(\hat{\pi}, \hat{\omega}$). 
Then, \eqref{+ep} is the same as \eqref{czachorp} with $\psi=\pi/2$ and overall sign positive. Similarly, \eqref{-ep} is the same as \eqref{czachorp} with $\psi=-\pi/2$ and overall sign negative.
 So we can inversely say that \eqref{czachorp} is the explicit expression for the real two-dimensional flag manifold $F_{2}$ of the double fibration\footnote{A similar form of solution has been known in the position twistor space, even though the physical meaning of the quantities obtained are quite different. See  \cite{penrosebook}, p58.}. 
Geometrically, the momenta $p^{(\pm)}_{\alpha\dot{\beta}}$ vary along two null geodesics with two fixed points $\pi^{0}_{\alpha} \bar{\pi}^{0}_{\dot{\beta}}$ and $-\pi^{0}_{\alpha} \bar{\pi}^{0}_{\dot{\beta}}$,
which are included in the null plane (called $\alpha$-plane) of the complex Minkowski space.

So far we have treated the general solution to the incidence  relation with the restrictions $\bar{Z}_A Z^A=0$ and ($\pi\cdot\omega)\neq 0$. And the solutions span the entire momentum Minkowski space. But, we have no prior reason to exclude the case  $(\pi\cdot\omega)=0$ with real momentum. If there exists any adequate solution with the condition, we can predict that the domain of the solution would be a subset of those of \eqref{+ep} and \eqref{-ep}.  The detailed analysis will be carried out in the next subsection, and will show  that the geomerical restriction ($\pi\cdot\omega)=0$ for nonzero $\pi$ and $\omega$ corresponds to the classical particle with imaginary mass, i.e., tachyon.

We can modify the same analysis we have given in this subsection to accommodate more general configurations of twistor coordinates, $\pi_\alpha \omega^\alpha = r^2 e^{i\psi}$. We may map $(\pi,\omega)$ to $(\pi', \omega')_\pm \equiv e^{-i(\psi/2\pm \pi/4)}(\pi,\omega)$ 
and declare that $(\pi,\omega)$ share the same solution sets as $(\pi',\omega')$. It is sensible since the solutions \eqref{+ep} and \eqref{-ep} are insensitive to the phases of $\pi$ or $\omega$. Alternatively, we may leave the twistor coordinates as they stand, and instead consider generalized hermitian form 
and incidence relations in the following way,  
\begin{align}
\bar{Z}_A Z^A|_\psi = e^{-i\psi} \pi_\alpha \omega^\alpha + e^{i\psi} \bar{\omega}^{\dot{\alpha}} \bar{\pi}_{\dot{\alpha}} \,,
\qquad
i e^{i\psi}p_{\alpha \dot{\beta}}\bar{\omega}^{\dot{\beta}} = \pi_{\alpha}\,.
\end{align}

To summarize, in the momentum twistor construction, a real massive momentum is expressed with one copy of twistor coordinates. This is different from the interpretation of massive momentum with position twistors in the sense that such an approach needs two copies of twistor sets \cite{hughston}.

\subsection{Momentum bispinor as the solution of the incidence relation with $\pi_{\alpha}\omega^\alpha=0$}

The condition $\pi_{\alpha}\omega^{\alpha}=0$ implies that $\pi_\alpha = k\omega_\alpha$  ($k \in \mathbb{C}$), so the incidence relation in this case can be rewritten as 
\begin{align}
k\omega_{\alpha}=ip_{\alpha\dot{\beta}}\bar{\omega}^{\dot{\beta}}.
\label{nincidence}
\end{align}
As in \eqref{generalp}, the general real momentum bispinor can be spanned by an auxiliary spinor $\lambda_{\alpha}$ that is independent of $\omega_{\alpha}$,  
\begin{align}
p_{\alpha\dot{\beta}} = 
c_1 \lambda_\alpha \bar{\lambda}_{\dot{\beta}} +
c_2 \lambda_\alpha \bar{\omega}_{\dot{\beta}} +
c_2 ^{*}\omega_\alpha \bar{\lambda}_{\dot{\beta}} +
c_4 \omega_\alpha \bar{\omega}_{\dot{\beta}},
\label{generalp2}
\end{align}
where $c_1$ and $c_4$ are real and $c_2$ is complex.
From \eqref{nincidence} ,
\begin{align}
k\omega_{\alpha}= i[c_{1}\lambda_{\alpha}(\bar{\lambda}\cdot\bar{\omega}) + c_{2}^{*}\omega_{\alpha}(\bar{\lambda}\cdot\bar{\omega})] \,,
\end{align}
we obtain $c_{1}=0$ and $k=ic_{2}^{*}(\bar{\lambda}\cdot\bar{\omega})$. Note that $c_2=0$ corresponds to $k=0$ and so $\pi_{\alpha}=0$. We can consider three different cases of the momentum form. 

First, when $c_{2}=0$ and $c_{4}\neq 0$, $\pi_\alpha$ vanishes and the momentum can be written as
\begin{align}
p_{\alpha\dot{\beta}} = c_4 \omega_\alpha \bar{\omega}_{\dot{\beta}} \,.
\end{align}
This is a massless momentum. 

Second,  when $c_{2}\neq 0$  and $c_{4}=0$, 
\begin{align}
p_{\alpha\dot{\beta}} = 
 \lambda_\alpha \bar{\omega}_{\dot{\beta}} +
\omega_\alpha \bar{\lambda}_{\dot{\beta}}
\end{align}
where $c_{2}$ is absorbed into redefined $\lambda_{\alpha}$. This momentum is tachyonic, i.e., $m^{2}=  -|\lambda\cdot \omega|^{2} =-|k|^2 <0$.

Third, when $c_{2}\neq 0$ and $c_{4}\neq 0$, 
\begin{align}
p_{\alpha\dot{\beta}} = 
 \lambda_\alpha \bar{\omega}_{\dot{\beta}} +
\omega_\alpha \bar{\lambda}_{\dot{\beta}} +
c_4 \omega_\alpha \bar{\omega}_{\dot{\beta}}\\
=c_{4}\Big(\omega_\alpha +\frac{1}{c_{4}}\lambda_\alpha\Big)\Big(\bar{\omega}_{\dot{\beta}}+\frac{1}{c_4}\bar{\lambda}_{\dot{\beta}}\Big) - \frac{1}{c_4}\lambda_\alpha \bar{\lambda}_{\dot{\beta}}\\
\equiv c_{4}\Omega_\alpha\bar{\Omega}_{\dot{\beta}} -\frac{1}{c_4}\lambda_\alpha \bar{\lambda}_{\dot{\beta}},
\end{align}
and the mass is tachyonic independent of $c_4$, $m^2=-|\lambda\cdot \omega|^{2}=-|k|^2 <0$.
 
The second and the third cases 
have little qualitatively difference in that, in both cases, $\pi_\alpha\neq 0$ and the mass is tachyonic. It is interesting to notice that the purely geometric condition $\pi_{\alpha}\omega^\alpha=0$ on the momentum twistor coordinate directly imposes a physical property on the corresponding classical particle, and $k$, the complex proportion of $\pi _\alpha$ and $\omega_\alpha$, determines the mass of particle as $i|k|$.


\subsection{Momentum/2-qubit correspondence revisited}

In the previous subsection, we showed that 
the momentum bispinor in Czachor's normalization \eqref{czachorp} 
can be interpreted as a solution to the incidence relation 
for the momentum twistor space. The new interpretation does 
not affect the identification of the momentum matrix 
as the density matrix in the 2-qubit entanglement system. 
A minor difference is that the concurrence $m=|\mathrm{det}C|$ is considered as a coordinate independent of the compactified projective space, which makes it possible to reach the $m=0$ (separable states) with $(\hat{\pi}\cdot \hat{\omega})$ remaining finite. 

There has been some effort to understand multipartite-qubit entanglement system with twistor coordinates. Especially, L\'evay \cite{levay2} gave a geometrical description of 3-qubit entanglement with SLOCC (stochastic local operation and classical communication) transformation (the equivalence of which is given by SL(2,$\mathbb{C}$) group) instead of LOCC transformation, and showed that the entangled states can be represented by a twistor space. In some sense, since we showed in section \ref{sec:f2} that the spinors that constitute a momentum bispinor can be understood as coordinates in the momentum twistor space, in a loose sense, we added another twistor-geometric approach to the entanglement system. But, it should be clarified that the manifold used here is not the projective space itself but the homogeneous coordinate of the space, even though the real scaling of the coordinate is projected effectively.

\section{Concluding remarks \label{sec:conclusion}}

We have analyzed the symmetry structure of massive momentum bispinors and applied the results to two geometric systems, 2-qubit entanglement and momentum twistor space. The correspondence between a region in Minkowski space and the 2-qubit entanglement system gives a simple and intuitive way to understand the 2-qubit entanglement system. Then, we investigated some geometric properties of the momentum twistor space as an attempt to understand the interesting projection property of Czachor's momentum bispinor, \eqref{p-incidence}. And in the setup we showed that a submanifold of the twistor space directly corresponds to the classical tachyon. We expect that our present research would be extended in two directions.

One is to investigate spinor sets (or twistor systems) in various dimensions and compare them with some entanglement systems. Since higher-dimensional spinors have more components, it is plausible that some of multipartite entangled qubit systems might be interpreted geometrically with higher dimensional momentum Minkowski space.
The other is to find a physical application aside from 
those concerning entangled qubits and classical tachyon. Our analysis of 
the momentum twistor space is so far remained geometric. Recall that the Penrose transform in the position twistor space relates solutions of massless field equations to holomorphic bundles on the twistor space \cite{penrose}. 
It would be interesting to explore a similar possibility for the momentum twistor space.

\vskip 1cm 

\subsection*{Acknowledgments}

The work of SL was supported in part by the National Research Foundation of Korea (NRF) Grants 2012R1A1B3001085 and 2012R1A2A2A02046739.

\vskip 2.5cm  


\centerline{\Large\bf Appendix}

\appendix

\section{Locus of det$A=0$ in $\mathbb{C}P^{3}$ \label{sec:app}}

In this section, we examine the locus det$A=0$ in two kinds of $\mathbb{C}P^{3}$ manifold in detail. 
{}From \eqref{sl2c}, we can see that SL(2,$\mathbb{C}$) = $\mathbb{C}P^{3} \setminus \textrm{(locus det}A'=0)$. So by examining  the locus det$A=0$ with the equivalence relation $A \sim A'$, we can obtain the exact manifold of SL(2,$\mathbb{C}$).  The general solution to det$A=0$ is given by
\begin{align}
A=
\begin{pmatrix} 
\pi_{1} & \pi_{2} 
\\
\omega_{1} & \omega_{2} 
\end{pmatrix}
=
\begin{pmatrix} 
\lambda_{1}\mu_{1} & \lambda_{1}\mu_{2} \\
\lambda_{2}\mu_{1} & \lambda_{2}\mu_{2}
\end{pmatrix}\,.
\end{align}
Consider the scaling by non-zero complex numbers,  
\begin{align}
(\lambda_1 , \lambda_2) \rightarrow \kappa_\lambda (\lambda_1 , \lambda_2) \,,
\quad 
(\mu_1 , \mu_2) \rightarrow \kappa_\mu (\mu_1 , \mu_2) \,,
\label{lm-gauge}
\end{align}
The overall scaling $\kappa_\lambda \propto \kappa_\mu$ 
is identified with the scaling that enters the definition of $\mathbb{C}P^{3}$. The relative scaling $\kappa_\lambda \propto 1/\kappa_\mu$ does not affect the values of $(\pi,\omega)$ 
and can be regarded as a gauge symmetry. To conclude, 
$(\lambda_1,\lambda_2)$ and $(\mu_1,\mu_2)$, with the equivalence relation with respect to the scalings  \eqref{lm-gauge}, parametrize $\mathbb{C}P^{1} \times \mathbb{C}P^{1} \subset \mathbb{C}P^{3}$. And the manifold that corresponds to SL(2,$\mathbb{C}$) is given by $\mathbb{C}P^{3} \setminus (\mathbb{C}P^{1} \times \mathbb{C}P^{1})$.

A similar result can be obtained in the context of twistor geometry discussed in section \ref{sec:twistor}. Here, $\mathbb{C}P^{3}$ manifold arises with a different equivalence relation, i.e., $(\pi_\alpha, \bar{\omega}^{\dot{\beta}}) \sim \eta(\pi_\alpha, \bar{\omega}^{\dot{\beta}}) $ ($\eta \in \mathbb{C}\setminus \{0\}$) but we still obtain the same $\mathbb{C}P^{1} \times \mathbb{C}P^{1}$ for the vanishing locus of det$A=0$ in twistor space. 
Finally, we note that the locus det$A=0$ is the same as 
the ``conifold" which plays an ubiquitous role in the geometry of string theory.

\end{document}